\documentclass[twocolumn]{aastex62}

\usepackage{comment}

\usepackage{graphicx}	% Including figure files
\usepackage{amsmath}	% Advanced maths commands
\usepackage{amssymb}	% Extra maths symbols
\usepackage[para,flushleft]{threeparttable} 
\usepackage{epstopdf}
\usepackage{textgreek}
\usepackage{comment}
\usepackage{multirow}

\usepackage{xcolor, fontawesome}
\definecolor{twitterblue}{RGB}{64,153,255}
\newcommand{\twitter}[1]{\href{https://twitter.com/#1}{\textcolor{twitterblue}{\faTwitter}\,\tt\hspace{2pt}\textcolor{blue!60!black}{@#1}}}

\hypersetup{colorlinks,linkcolor={red!60!black},citecolor={blue!60!black},urlcolor={blue!60!black}}

\usepackage{natbib,twoopt}
\bibpunct{(}{)}{;}{a}{}{,} %% natbib format for A&A and ApJ
\makeatletter
\newcommandtwoopt{\citeads}[3][][]{\href{http://ui.adsabs.harvard.edu/\#abs/#3}%
{\def\hyper@linkstart##1##2{}%
\let\hyper@linkend\@empty\citealp[#1][#2]{#3}}}
\newcommandtwoopt{\citepads}[3][][]{\href{http://ui.adsabs.harvard.edu/\#abs/#3}%
{\def\hyper@linkstart##1##2{}%
\let\hyper@linkend\@empty\citep[#1][#2]{#3}}}
\newcommandtwoopt{\citetads}[3][][]{\href{http://ui.adsabs.harvard.edu/\#abs/#3}%
{\def\hyper@linkstart##1##2{}%
\let\hyper@linkend\@empty\citet[#1][#2]{#3}}}
\newcommandtwoopt{\citeyearads}[3][][]%
{\href{http://ui.adsabs.harvard.edu/\#abs/#3}
{\def\hyper@linkstart##1##2{}%
\let\hyper@linkend\@empty\citeyear[#1][#2]{#3}}}
\makeatother

\graphicspath{{./}{figures/}}

\accepted{\today}
\submitjournal{ApJL}

\shorttitle{An Earth-like stellar environment for Proxima Centauri c} 
\shortauthors{Alvarado-G\'omez \& Drake et al.}
\begin{document}

\title{An Earth-like stellar wind environment for Proxima Centauri c}

\correspondingauthor{J. D. Alvarado-G\'omez \& J. J. Drake\\(These authors contributed equally to this work).}
\email{julian.alvarado-gomez@aip.de, jdrake@cfa.harvard.edu}

\author[0000-0001-5052-3473]{Juli\'an D. Alvarado-G\'omez}
\altaffiliation{\twitter{AstroRaikoh} $|$ Karl Schwarzschild Fellow}
\affil{Leibniz Institute for Astrophysics Potsdam, An der Sternwarte 16, 14482 Potsdam, Germany}

\author[0000-0002-0210-2276]{Jeremy~J.~Drake}
\altaffiliation{\twitter{cosmodrake}}
\affil{Center for Astrophysics $|$ Harvard \& Smithsonian, 60 Garden Street, Cambridge, MA 02138, USA}
\collaboration{}

\author[0000-0002-8791-6286]{Cecilia Garraffo}
\affil{Institute for Applied Computational Science, Harvard University, Cambridge, MA 02138, USA}
\affil{Center for Astrophysics $|$ Harvard \& Smithsonian, 60 Garden Street, Cambridge, MA 02138, USA}

\author[0000-0003-3721-0215]{Ofer~Cohen}
\affil{University of Massachusetts at Lowell, Department of Physics \& Applied Physics, 600 Suffolk Street, Lowell, MA 01854, USA}

\author[0000-0003-1231-2194]{Katja Poppenhaeger}
\affil{Leibniz Institute for Astrophysics Potsdam, An der Sternwarte 16, 14482 Potsdam, Germany}
\affil{University of Potsdam, Institute for Physics and Astronomy, Karl-Liebknecht-Str. 24/25, 14476 Potsdam, Germany}

\author[0000-0002-9569-2438]{Rakesh K. Yadav}
\affiliation{Department of Earth and Planetary Sciences, Harvard University, Cambridge, MA 02138, USA}

\author[0000-0002-2470-2109]{Sofia~P.~Moschou}
\affil{Center for Astrophysics $|$ Harvard \& Smithsonian, 60 Garden Street, Cambridge, MA 02138, USA}



\begin{abstract}
\noindent A new planet has been recently discovered around Proxima Centauri. With an orbital separation of $\sim$\,$1.44$~au and a minimum mass of about $7~M_{\earth}$, Proxima~c is a prime direct imaging target for atmospheric characterization. The latter can only be performed with a good understanding of the space environment of the planet, as multiple processes can have profound effects on the atmospheric structure and evolution. Here, we take one step in this direction by generating physically-realistic numerical simulations of Proxima's stellar wind, coupled to a magnetosphere and ionosphere model around Proxima~c. We evaluate their expected variation due to the magnetic cycle of the host star, as well as for plausible inclination angles for the exoplanet orbit. Our results indicate stellar wind dynamic pressures comparable to present-day Earth, with a slight increase (by a factor of 2) during high activity periods of the star. A relatively weak  interplanetary magnetic field at the distance of Proxima~c leads to negligible stellar wind Joule heating of the upper atmosphere (about $10\%$ of the solar wind contribution on Earth) for an Earth-like planetary magnetic field ($0.3$~G). Finally, we provide an assessment of the likely extreme conditions experienced by the exoplanet candidate Proxima~d, tentatively located at $0.029$~au with a minimum mass of $0.29$~$M_{\earth}$.   
\end{abstract}



%@arxiver{Fig1_ProxCen_B.jpg,GM_High-Activity.jpeg}

\section{Introduction} \label{sec:intro}

\noindent The space weather in M dwarf planetary systems presents a particularly challenging case for exoplanet atmospheres. The diminutive bolometric luminosity ($L_{\rm bol}$) of M dwarfs means their temperature-based habitable zones, within which liquid water can be sustained, lie close to the central star---as much as ten or more times closer than in the case of our own solar system (e.g.~Kopparapu~et al.~\citeyearads{2013ApJ...770...82K}, \citeyearads{2014ApJ...787L..29K} \citeads{2016PhR...663....1S}). 

Specifically, the habitable zone semi-major axis ($a$) follows $a^2 \propto L_{\rm bol}$ \citepads{2013ApJ...770...82K}. On the other hand, stellar wind mass loss rates are thought to scale with X-ray luminosity with a power greater than one (i.e.,~$\dot{M}_{\bigstar} \propto L^{1.34}_{\rm X}$, \citeads{2005ApJ...628L.143W}). In this way, the wind through a unit surface area at the main sequence habitable zone distance scales like $\dot{M}_{\bigstar}/a^2 \propto L^{1.34}_{\rm X}/L_{\rm bol} \propto L^{1.34}_{\rm X}/M_{\bigstar}^{3.5}$. Therefore, the stellar wind intensity within the habitable zone increases with decreasing stellar mass. Moreover, M dwarfs remain magnetically very active (i.e., high $L_{\rm X}/L_{\rm bol}$ values) over much longer timescales than higher mass stars (e.g.~\citeads{2011ApJ...743...48W}, \citeads{2019ApJ...871..241D}), so that the integrated exposure to the most intense stellar winds is commensurately greater.

A number of studies employing detailed and realistic magnetohydrodynamic (MHD) simulations of stellar winds have examined the effects of space weather on exoplanets, including those in M dwarf systems (e.g.~\citeads{2014MNRAS.438.1162V}, \citeads{2014ApJ...790...57C}, \citeads{2019ApJ...875L..12A}). For habitable zone planets around M dwarfs, models predict stellar wind dynamic pressures up to four orders of magnitude greater than experienced by the present-day Earth, together with orders of magnitude pressure variations on sub-orbital timescales of one to a few days (\citeads{2014MNRAS.438.1162V}, Garraffo et~al.~\citeyearads{2016ApJ...833L...4G},  \citeyearads{2017ApJ...843L..33G}), intense Joule heating (Cohen~et~al.~\citeyearads{2014ApJ...790...57C}, \citeyearads{2018ApJ...856L..11C}), severe atmospheric loss (\citeads{2017ApJ...837L..26D}, \citeads{2017ApJ...844L..13G}), and transitions into and out of sub-Alfv\'enic wind conditions on orbital timescales (\citeads{2014ApJ...790...57C}, \citeads{2017ApJ...843L..33G}). 

Here, we study the steady stellar wind environment of the new-found planetary companion around our nearest star (Proxima~c, \citeads{2020SciA....6.7467D}). The discovery of a planetary system around Proxima \citepads{2016Natur.536..437A} represented a watershed moment in exoplanetary research---a stark confirmation that planetary systems are very common in the universe with the tantalizing prospect of potentially being reachable by an interstellar probe (e.g.~\citeads{2017AJ....154..115H}, \citeads{2018AcAau.152..370P}). We construct three-dimensional MHD models of the magnetized stellar wind of Proxima using a state-of-the-art computational framework and a surface magnetic field map derived from sophisticated dynamo simulations tuned to the case of Proxima \citepads{2016ApJ...833L..28Y}. 

We use the models to investigate the conditions experienced by Proxima c, which is estimated to have a mass of about seven times that of Earth, orbiting at a distance comparable to Mars in the solar system \citepads{2020SciA....6.7467D}. We evaluate how the stellar wind properties change with the magnetic activity level of the host star, and compare our results to those of previous studies on other exoplanet systems. Finally, we touch upon the case of Proxima d, a tentative additional planetary candidate of the Proxima system that would be the closest known planet to the star, lying within the orbit of Proxima b \citepads{2020A&A...639A..77S}. 

\section{The Proxima Centauri System} \label{s:proxima}

\noindent Proxima Centauri, also just known as Proxima, is an M5.5 dwarf with an effective temperature of $3042$~K, a mass of $0.122~M_\odot$, a radius of $0.154~R_\odot$, a rotation period of $83$~days, and an estimated age of $4.85$~Gyr (\citeads{2003A&A...397L...5S}, \citeads{2007AcA....57..149K}, \citeads{2016Natur.536..437A}).

Proxima hosts our nearest exoplanetary system and presents a unique opportunity for exoplanet characterization. The first planet discovered in the system, Proxima b, is estimated to be of at least $1.17$~Earth masses (\citeads{2020A&A...639A..77S}) and has an orbital period of $11.2$~days, with a semi-major axis of only $0.049$~au \citepads{2016Natur.536..437A}. This orbit is approximately twenty times closer to Proxima than the Earth is to the Sun.  
Proxima b does not transit Proxima Centauri from the vantage point of the solar system \citepads{2019MNRAS.487..268J} and its orbital inclination, and consequently its mass, are presently unknown. 

Proxima b is in Proxima's classically-defined ``habitable zone'', having an equilibrium temperature of $234$~K \citepads{2016Natur.536..437A}, which is slightly cooler than that of Earth ($255$~K). Several studies have examined its likely irradiation history and possible climate and evolution in relation to potential habitability (e.g.~\citeads{2016A&A...596A.111R}, \citeads{2016A&A...596A.112T}).

Analysis of radial velocity variations by \citetads{2020SciA....6.7467D} suggested the presence of a secondary $\sim6-7~M_\Earth$ planet in a $\sim5$~yr orbit around Proxima. Follow-up studies have placed limits on the properties of Proxima~c, measuring anomalies in Proxima's astrometric proper motion (\citeads{2020A&A...635L..14K}, \citeads{2020RNAAS...4...46B}), as well as direct imaging from ground-based observations \citepads{2020A&A...638A.120G}. Combining all the available constraints, \citetads{2020RNAAS...4...86B} obtained the most up-to-date set of orbital parameters, placing it in a circular orbit ($e\simeq0$) at approximately $1.44$~au ($\sim$\,$5.3$~yr orbital period).  

A recent study reports a small radial velocity perturbation of Proxima which, assuming a planetary origin, would indicate an additional low-mass object ($M\sin i\simeq0.29~M_\Earth$) at a distance of $\sim0.029$~au (\citeads{2020A&A...639A..77S}). If confirmed, it would become the innermost known planet of the system, orbiting closer than Proxima b and just short of the optimistic habitable zone ($\sim0.03-0.09$~au, \citeads{2012PASP..124..323K}).

Harsh circumstellar conditions are expected in the system. Proxima itself is a flare star and displays optical, UV and X-ray variability that is consistent with a stellar activity cycle with a period of about 7 years \citepads{2017MNRAS.464.3281W}. The amplitude of the cycle in the stellar X-ray luminosity is of the order of $\pm 50\%$ in an energy band of 1.2-2.4~keV, and somewhat lower in softer X-rays at energies 0.2-1.2~keV at approximately $\pm 20\%$. This means that apart from flares, which occur often on Proxima (e.g.~\citeads{2011A&A...534A.133F}, \citeads{2019ApJ...884..160V}), the quiescent X-ray environment of the planets also changes over time. 

Unlike the coronal properties, only limits are available on the steady and transient outflows from Proxima. Observations of the Ly-$\alpha$ astrospheric absorption indicate a stellar wind mass loss rate\footnote[7]{Assuming $\dot{\rm M}_{\odot} \simeq 2 \times 10^{-14}$~M$_{\odot}$ yr$^{-1}$ $= 1.265 \times 10^{12}$~g~s$^{-1}$} $\dot{\rm M}_{\bigstar} < 0.2~\dot{\rm M}_{\odot}$ \citepads{2001ApJ...547L..49W}, while wind-ISM charge exchange X-ray signatures place it at $\dot{\rm M}_{\bigstar} < 14~\dot{\rm M}_{\odot}$ \citepads{2002ApJ...578..503W}. Likewise, despite its frequent flaring, there are no direct detections of coronal mass ejections in Proxima so far (see~\citeads{2019ApJ...877..105M} and references therein). 

\begin{figure*}[!ht]
\centering%  left, bottom, right and top
\includegraphics[trim=0.15cm 0.0cm 0.15cm 1.0cm, clip=true,scale=0.1475]{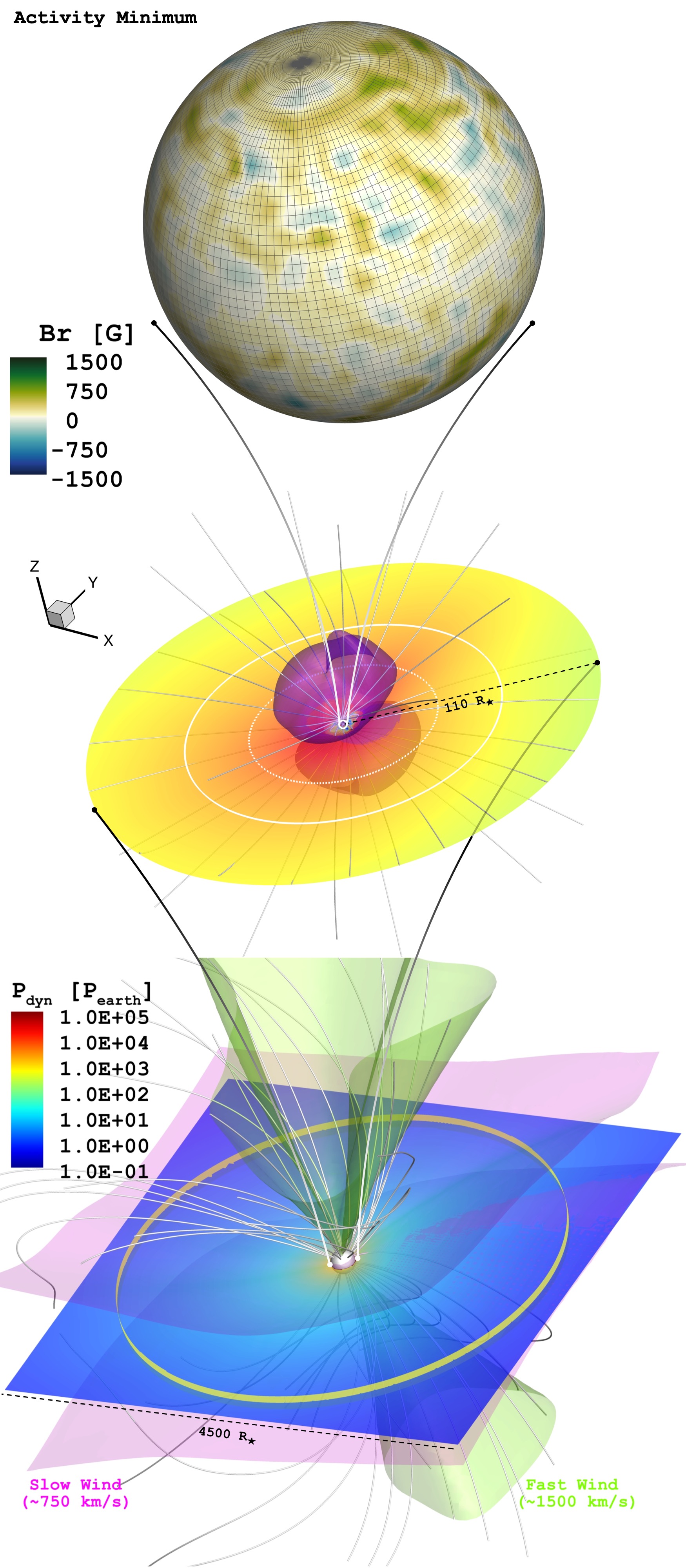}\includegraphics[trim=0.15cm 0.0cm 0.15cm 1.0cm, clip=true,scale=0.1475]{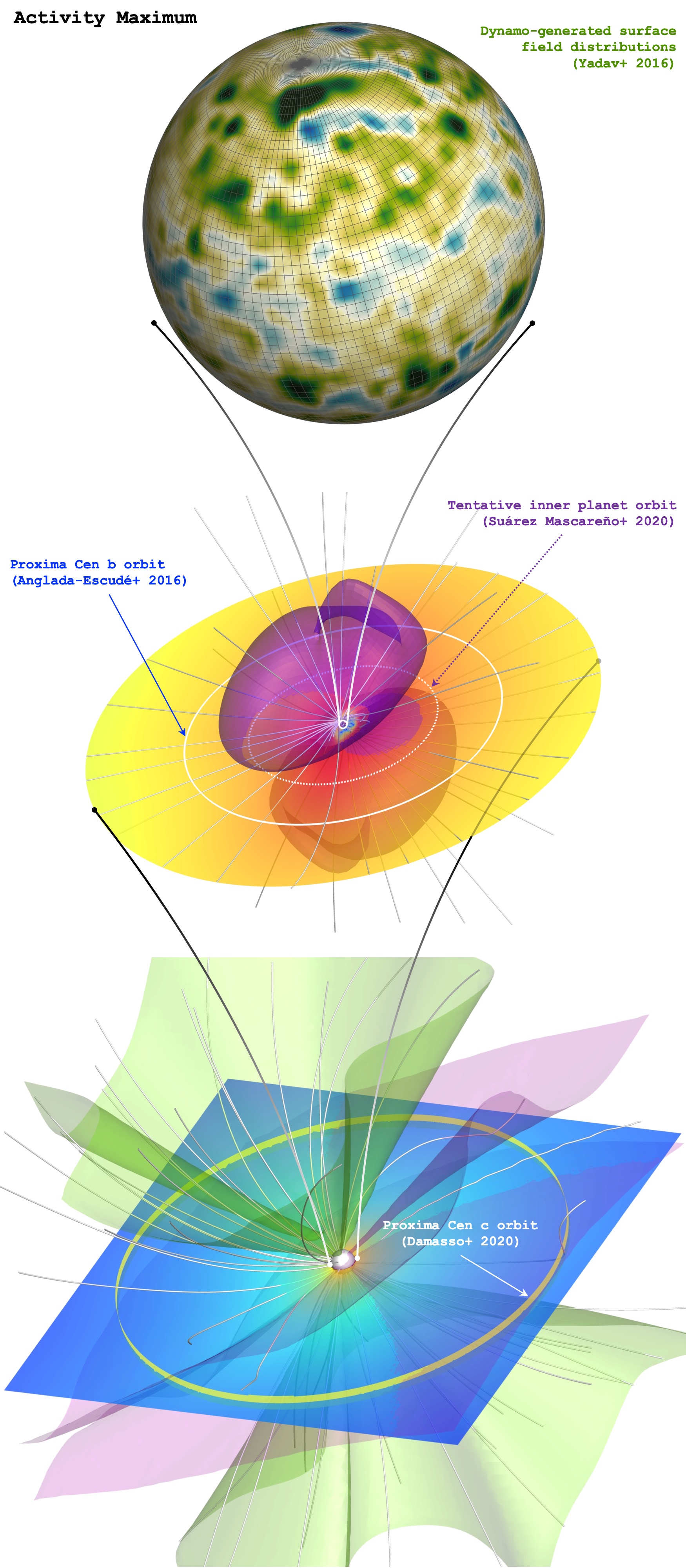}\vspace{-0.15cm}
\caption{Simulated stellar wind environment for the Proxima Cen system. Multi-domain models for activity minimum (left) and maximum (right) are shown. The top panels contain the dynamo-generated surface field distributions (in G) used to drive the AWSoM solution within the innermost module (SC, middle panels). This domain contains the orbits of Proxima b (white solid) and the tentative innermost planet Proxima d (white dashed). The purple iso-surface corresponds to the Alfv\'en surface of the stellar wind ($M_{\rm A} = 1$, see text for details). The steady-state solution is propagated from the coupling region ($105-110$~$R_{\bigstar}$) to the entire IH domain ($4500$~$R_{\bigstar}$ in each cartesian direction; bottom panels). This domain contains the orbit of Proxima~c (yellow). Magenta and green iso-surfaces delimitate the slow ($U_{\rm r} \lesssim 750$ km/s) and fast ($U_{\rm r} \gtrsim 1500$ km/s) wind sectors, respectively.  Color-coded is the wind dynamic pressure ($P_{\rm dyn} = \rho\,U^2$) normalized to the nominal Sun-Earth value ($\sim$\,$1.5$~nPa), visualized on the equatorial plane of both domains. Selected magnetic field lines are shown in white.}
\label{fig_1}
\end{figure*}

\section{Numerical Methodology} \label{sec:Methods}

\noindent Our characterization of the stellar wind conditions in the Proxima Cen system employs the state-of-the-art Space Weather Modeling Framework (SWMF, \citeads{2018LRSP...15....4G}). Originally developed for solar system studies, the SWMF contains a collection of physics-based models that can be executed individually or can be coupled to cover a wide range of regions within the space environment of the Sun (e.g., from the convection zone to the outer heliosphere; see \citeads{2012JCoPh.231..870T}). The simulations presented here consider four modules of the SWMF, covering the stellar corona (SC, $\sim$\,$1.0 - 110~R_{\bigstar}$), the inner heliosphere\footnote[8]{Denoted as inner astrosphere in the stellar case.} (IH, $105 - 2250~R_{\bigstar}$), the global magnetosphere (GM\footnote[9]{The domain size is defined in units of planetary radii instead of stellar radii in this case.}, day side: $100~R_{\rm p}$, night side: $225~R_{\rm p}$, North-South: $256~R_{\rm p}$), and a domain for ionospheric electrodynamics (IE\footnote[10]{A two-dimensional sphere set at an altitude of 120 km in the case of the Earth.}).  

The multi-domain solution is constructed from inside out, initially calculated within the SC module using the Alfv\'en Wave Solar Model (AWSoM, \citeads{2014ApJ...782...81V}), whose standard boundary conditions are modified to the M-dwarf regime (and specifically for Proxima) as described in \citetads{2020ApJ...895...47A}. In particular, two surface magnetic field configurations ---associated with minimum and maximum activity--- are considered to drive individual AWSoM solutions. These have been extracted\footnote[11]{Rotations $508$ (minimum) and $520$ (maximum) from the \citetads{2016ApJ...833L..28Y} study are used for this purpose.} from a self-consistent fully-convective dynamo simulation ---adjusted to the stellar mass, radius, and rotation period of Proxima (see~\citeads{2016ApJ...833L..28Y})--- whose oscillatory regime yields a time-scale comparable with the observed activity cycle in the star (Sect.~\ref{s:proxima}). As shown in the top panels of Fig.~\ref{fig_1}, we scale the average surface field strengths for activity minimum ($450$~G) and maximum ($750$~G) to match the limits from Zeeman broadening observations of Proxima ($\left<B\right>_{\rm S} = 600 \pm 150$~G; \citeads{2008A&A...489L..45R}).  

Once an AWSoM steady-state is achieved, it gets propagated via three different couplings with the other domains. The first one connects the outer boundary of SC with the inner boundary of IH (with a $5~R_{\bigstar}$ domain overlap), while the second one is performed within the IH domain (along the specified orbit of Proxima c) establishing the upstream stellar wind conditions as one of the outer (side) boundaries of GM. As the magnetosphere relaxes, the third coupling takes place, where field-aligned currents are computed in GM, and then are mapped to the IE domain assuming a planetary dipole field of $-0.3$~G (aligned with $z$-axis) in the case of Proxima~c. We stress here that there are no observational constraints on the magnetization of this exoplanet.  The planetary field selection was made to ease the comparison with the case of the Earth and other systems studied with a similar methodology (e.g.~\citeads{2019ApJ...875L..12A}, \citeads{2020ApJ...897..101C}). 
The IE module uses the field aligned currents to calculate the flux of the precipitating electrons, and the energy dissipating in the ionosphere (Joule Heating, hereafter JH) assuming a specific conductance pattern that can be either a constant Pedersen conductivity, or a more complicated conductivity pattern that can be obtained from other models or data. In the Earth case, the integrated conductivity ranges between $1$~S and $10$~S, where a lower value leads to an increased JH \citepads{2014ApJ...790...57C}. For simplicity, we use a constant conductivity of $1$~S to estimate an upper limit to the JH under the assumed stellar wind parameters. The IE provides an improved boundary conditions for GM in the form of electric and velocity fields at the inner boundary. We refer the reader to \citeads{2020ApJ...897..101C} for more details about the GM-IE coupling, and the JH calculation.

A combination of spherical (SC) and Cartesian (IH/GM) grids is employed, which is further optimized using multiple realizations of adaptive mesh refinement/coarsening, informed by magnetic field and particle density gradients. This was necessary to keep the number of cell blocks tractable, given the very large IH box size (side~length:~$4500~R_{\bigstar}$) required to contain the complete orbit of Proxima~c ($a\simeq~2010.39~R_{\bigstar}$, $e = 0.0$; see \citeads{2020SciA....6.7467D}, \citeads{2020RNAAS...4...86B}). In this way, the combined domain contains more than $24$ million spatial blocks, with the smallest cell elements in the final mesh reaching $0.025~R_{\bigstar}$ (SC), $4.394~R_{\bigstar}$ (IH), and $0.3~R_{\rm p}$~(GM).% in their respective domain units. 

\newpage
\section{Results and Discussion} \label{sec:results}

\noindent Results from our numerical simulations of the Proxima system are presented in Figs.~\ref{fig_1} to \ref{fig_GM}, where side-by-side visualizations for activity minimum and maximum are shown. As described below, good agreement is obtained with current observational constraints on Proxima's stellar wind, as well as with previous modeling work by \citetads{2016ApJ...833L...4G} on the space weather conditions around Proxima~b (Fig.~\ref{fig_1}, middle~panels).

\subsection{Stellar Wind Models}\label{sec:wind}

\noindent Despite the differences in surface field strength and topology (Fig.~\ref{fig_1}, top panels), the resulting steady-state wind solutions are similar between both activity states. This is a consequence of comparable large-scale magnetic field components among both configurations, with the small-scale structure mostly controlling the coronal thermodynamic conditions (see \citeads{2015ApJ...798..116R}, \citeads{2015ApJ...807L...6G}). The associated Alfv\'en Surface (AS)\footnote[12]{Defined by the locations in which the stellar wind speed matches the local Alfv\'en speed (i.e., an Alfv\'enic Mach number $M_{\rm A} = U\sqrt{4\pi\rho}/B = 1$, where $U$, $\rho$ and $B$, correspond to the wind speed, density and magnetic field values, respectively).} displays a characteristic two-lobe configuration (Fig.~\ref{fig_1}, middle panels), with average sizes of $28.1~R_{\bigstar}$ and $46.5~R_{\bigstar}$ for activity minimum and maximum, respectively. The wind distribution is mainly bipolar (see~Fig.~\ref{fig_1}, bottom panels), with a relatively fast component reaching up to $\sim$\,$1500$~km~s$^{-1}$ in the (magnetic) pole-ward directions, and a slow wind sector ($\lesssim$\,$750$~km~s$^{-1}$) surrounding the astrospheric current sheet. The latter is roughly aligned with the equatorial plane during minimum, gaining a small inclination angle ($\sim$\,$20^{\circ}$) for activity maximum. 

Computing the stellar wind mass loss rate for each magnetic configuration yields $\sim$\,$0.3$~$\dot{\rm M}_{\odot}$ (minimum) and $\sim$\,$0.9$~$\dot{\rm M}_{\odot}$ (maximum). These values appear close to current upper limits from observations (see~Sect.~\ref{s:proxima}). Note also that the factor of $3$ difference in $\dot{\rm M}_{\bigstar}$ between activity states is comparable to the observed variation in $\dot{\rm M}_{\odot}$ over the solar cycle (by a factor of $\sim$\,$2$; \citeads{2018ApJ...864..125F}).

\subsection{Stellar Wind Environment of Proxima~c}\label{sec:Prox-c}

\noindent Having established that our simulations provide a robust description of Proxima's stellar wind, we now proceed to assess the expected conditions for planet c. For each activity state, the bottom panels of Fig.~\ref{fig_1} display the resulting stellar wind dynamic pressure, $P_{\rm dyn} = \rho\,U^2$ (normalized to the average value experienced by the Earth, $P_{\rm earth} \simeq 1.5$~nPa\footnote[13]{\href{https://www.swpc.noaa.gov/products/real-time-solar-wind}{https://www.swpc.noaa.gov/products/real-time-solar-wind}}), up to the orbital distance of Proxima c. 

\begin{figure*}[t] %  left, bottom, right and top
\includegraphics[trim=0.6cm 0.75cm 0.3cm 4.2cm, clip=true,height=0.292\textwidth]{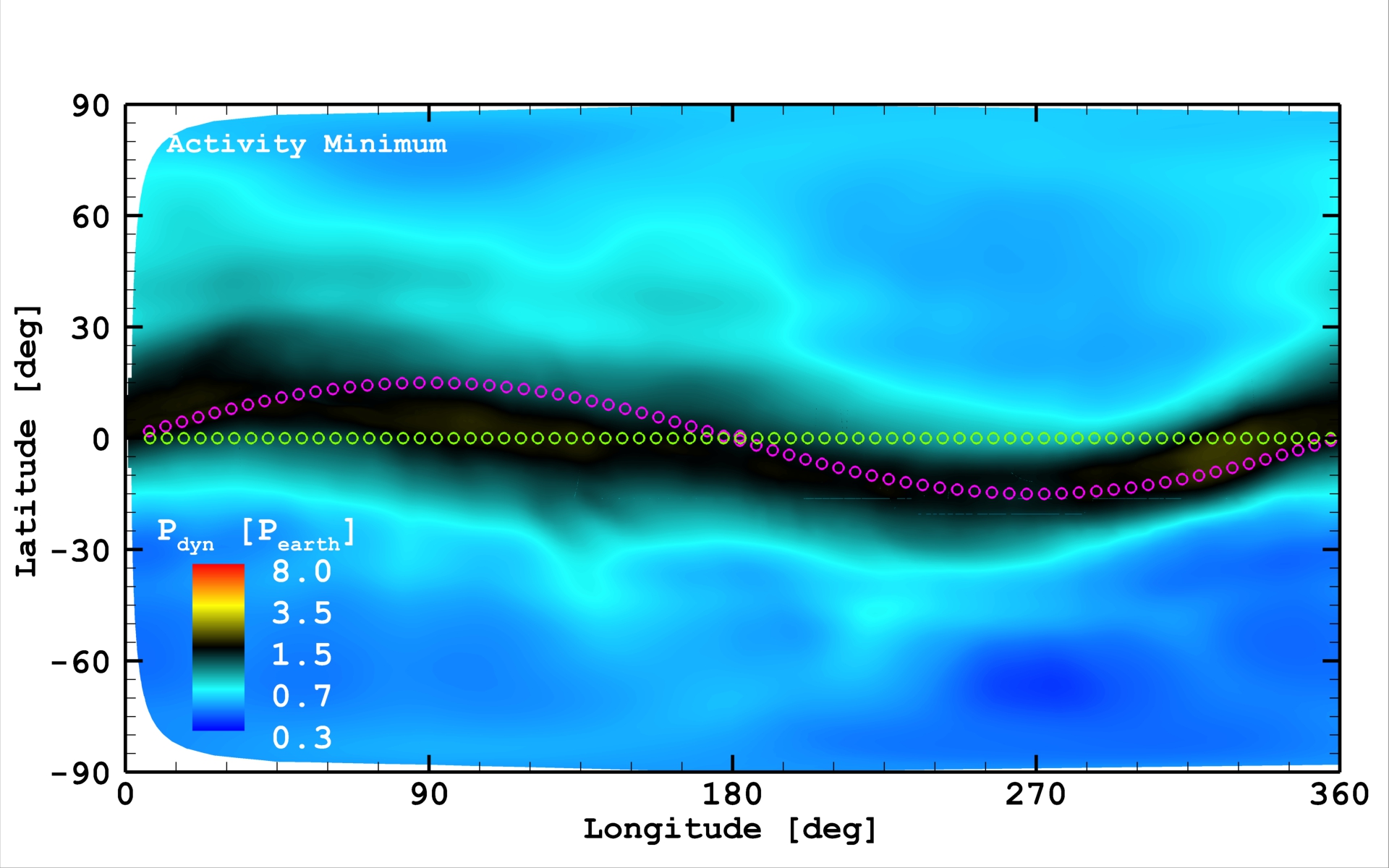}\includegraphics[trim=5.5cm 0.75cm 0.3cm 4.2cm, clip=true,height=0.292\textwidth]{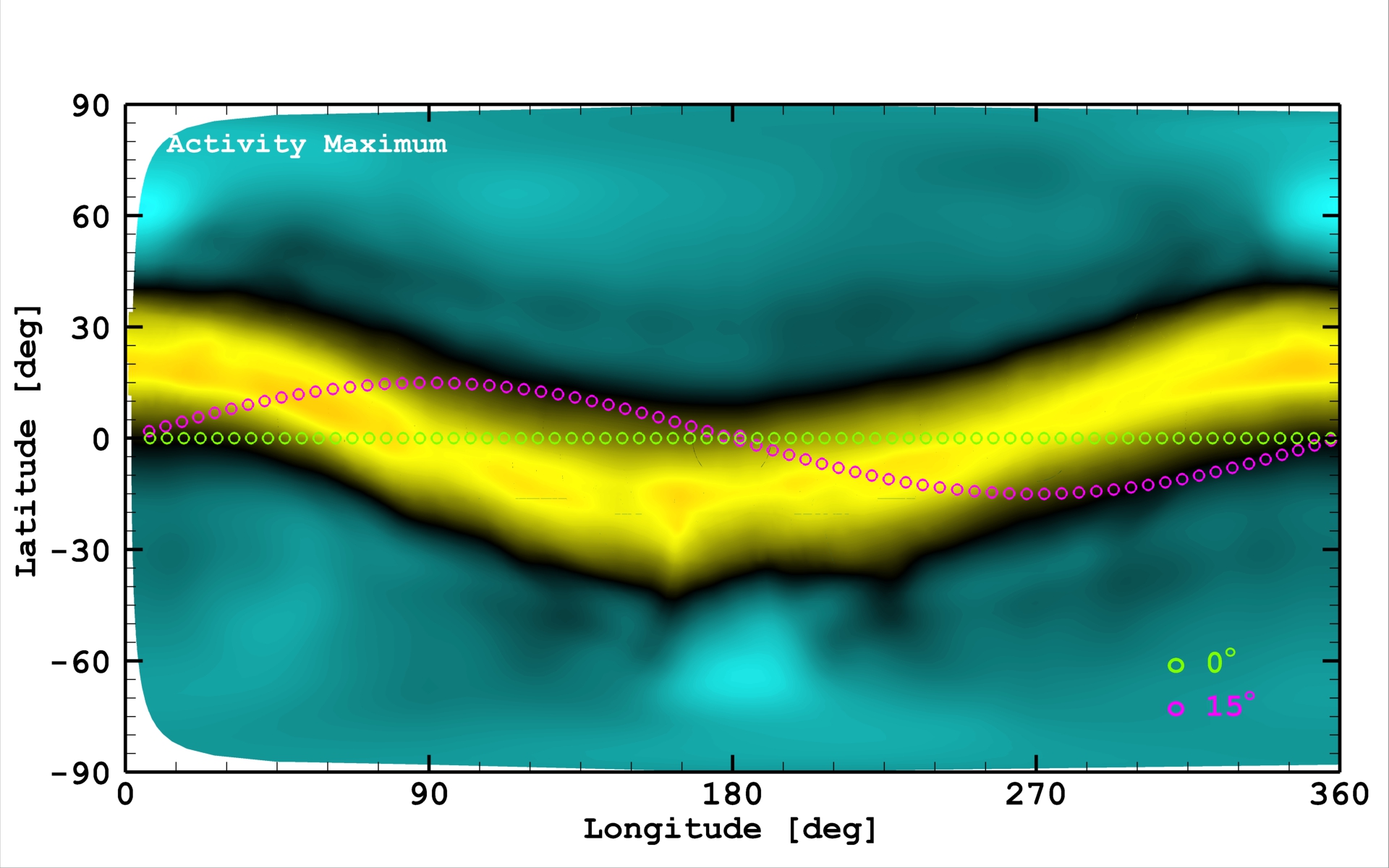}
\caption{Two-dimensional Mercator projection of the normalized stellar wind dynamic pressure ($P_{\rm dyn}$), extracted from a sphere at the distance of Proxima~c ($\sim\,1.44$~au $\simeq 2010.39~R_{\bigstar}$). Black and purple dotted lines indicate the path for $0^{\circ}$ and $15^{\circ}$ orbital inclinations, respectively. Conditions for activity minimum (left) and maximum (right) are shown.}
\label{fig:Pdyn_2D}
\end{figure*}

To examine their expected orbital variations, Fig.~\ref{fig:Pdyn_2D} shows two-dimensional Mercator projections of $P_{\rm dyn}$ constructed from a sphere with radius matching the semi-major axis of Proxima c ($\sim$\,1.44~au). We include the orbital paths for two possible inclinations\footnote[14]{Measured with respect to the equatorial plane and not with respect to the line-of-sight (which is the value reported in \citeads{2020A&A...635L..14K} and \citeads{2020RNAAS...4...86B}).} of the planet ($0^{\circ}$ and $15^{\circ}$). During minimum, the largest value in $P_{\rm dyn}$ along the explored orbits is close to two times the Sun-Earth average. The conditions worsen slightly for activity maximum, with a stellar wind dynamic pressure reaching up to $4~P_{\rm earth}$. 

\begin{figure*}
\center %  left, bottom, right, top
\includegraphics[trim = 0.2cm 0.25cm 0.25cm 0.cm, clip=true,scale=0.445]{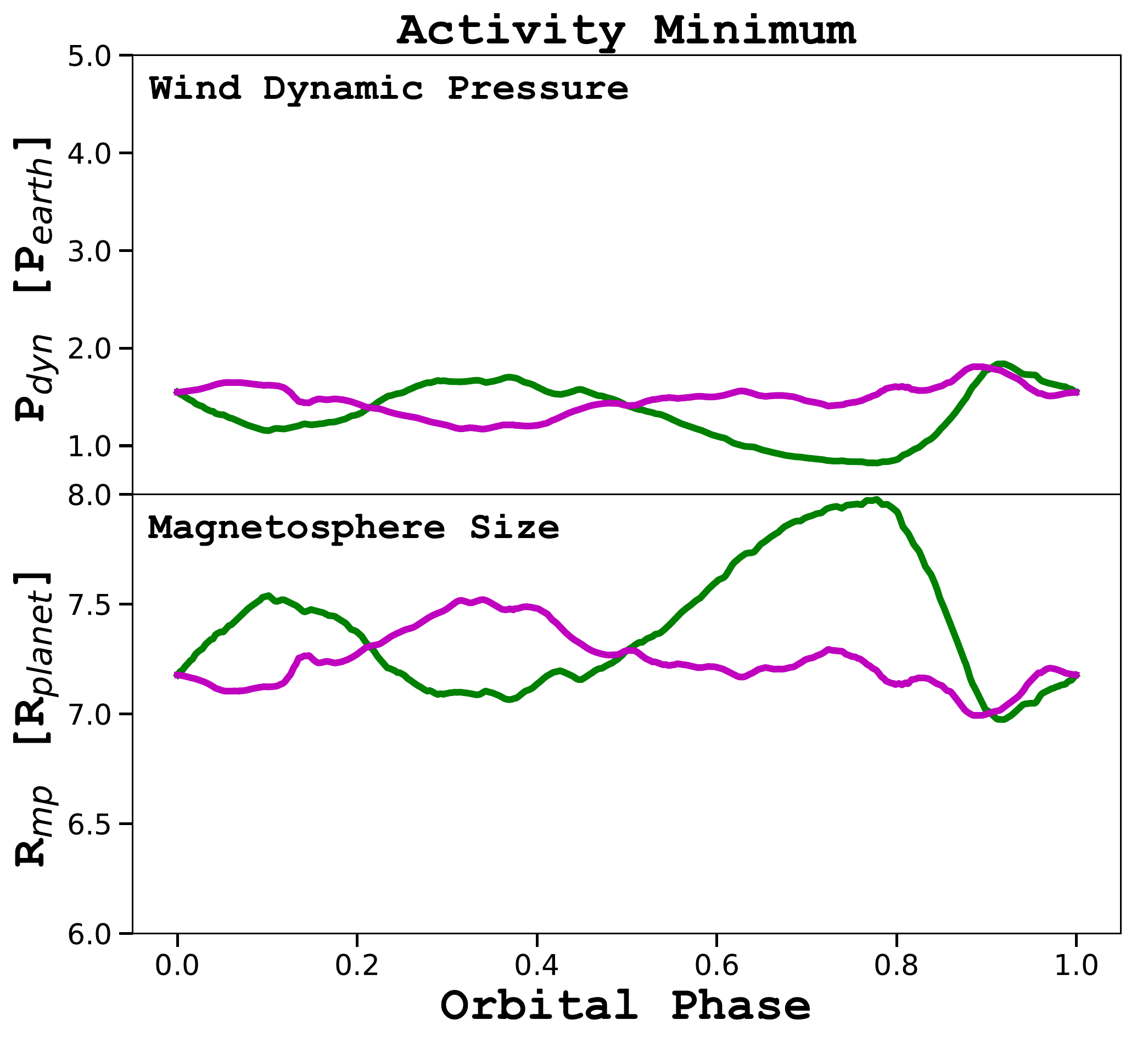}\includegraphics[trim = 2.2cm 0.25cm -0.3cm 0.0cm, clip=true,scale=0.445]{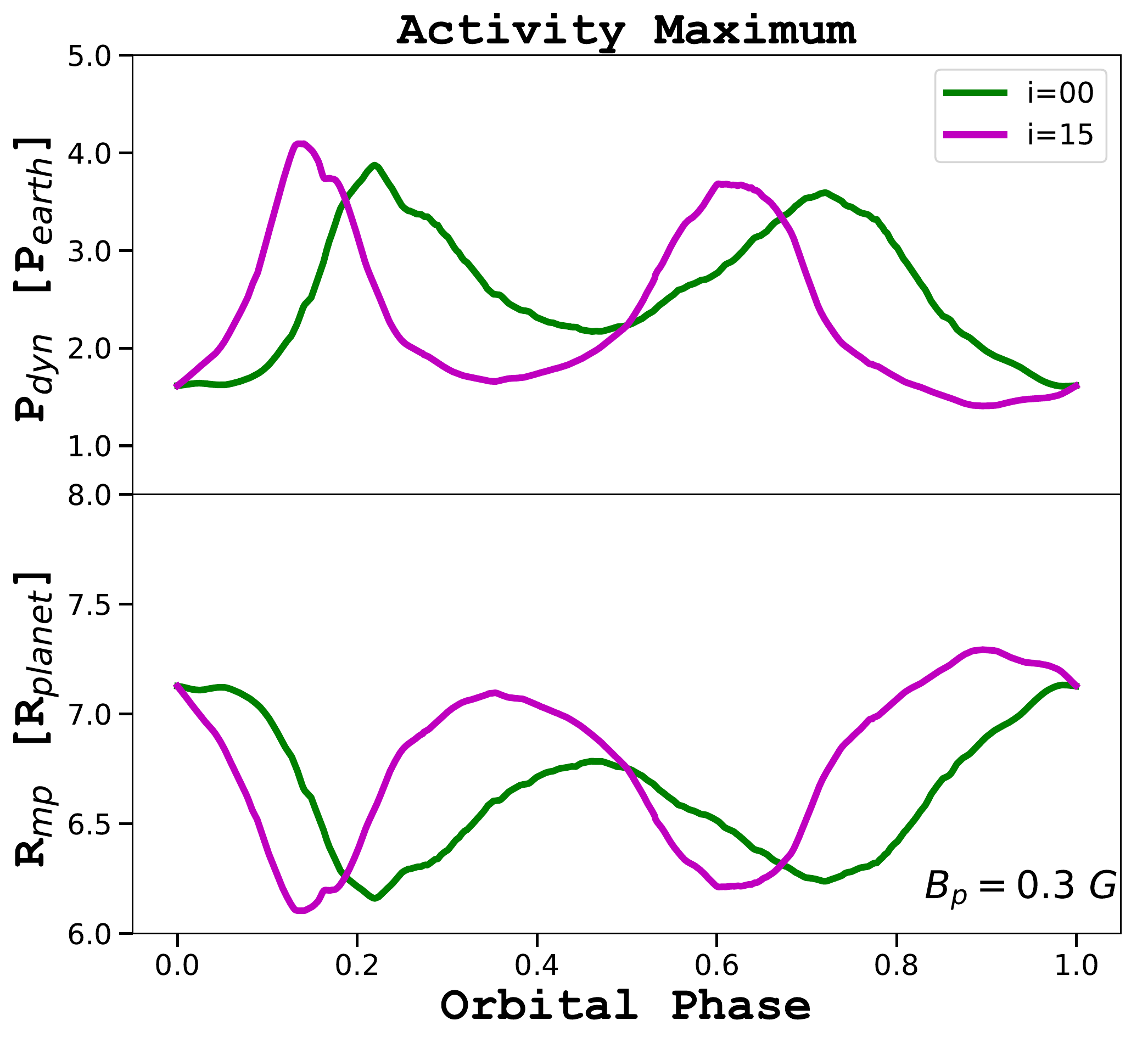}
\caption{Variation of the normalized wind dynamic pressure ($P_{\rm dyn}$, top) and the magnetosphere size ($R_{\rm mp}$, bottom) along possible orbits of Proxima c. Black and purple lines show the behavior for orbital inclinations of $0^{\circ}$ and $15^{\circ}$, respectively (see~Fig.~\ref{fig:Pdyn_2D}). Conditions for both stellar activity states are included (minimum: left, maximum: right). An Earth-like planetary dipole magnetic field ($B_{\rm p} = 0.3$~G) is assumed in all cases.}
\label{fig:orbits}
\end{figure*}

For the considered inclinations, the orbital variability of $P_{\rm dyn}$ is rather small, being around $50\%$ for activity minimum and close to a factor of $2$ during maximum. This is better illustrated in the top panels of Fig.~\ref{fig:orbits}, showing the behaviour of $P_{\rm dyn}$ as a function of orbital phase in all cases. With a more inclined orbit, Proxima c would be exposed to stellar wind sectors of substantially lower dynamic pressure, at the cost of enhanced variability during each current sheet crossing (e.g.,~\citeads{2016A&A...594A..95A}, \citeads{2016ApJ...833L...4G}). 

Interestingly, our $P_{\rm dyn}$ results for Proxima~c are comparable with expectations for Barnard Star~b (see~\citeads{2019ApJ...875L..12A}), which resides much closer to its host star ($a \simeq 0.4$~au, $e \simeq 0.32$, \citeads{2018Natur.563..365R}). At a considerable older age ($\sim$\,$10$~Gyr), the weaker magnetism of Barnard Star creates a slower and more rarefied stellar wind compared to Proxima. This compensates the shorter orbital distance, leading to similar $P_{\rm dyn}$ conditions for both super-Earth planets.

The super-Alfv\'enic stellar wind conditions along the orbit, combined with the assumption of a dipole planetary magnetic field ($B_{\rm p}$), allow a broad estimate on the associated day-size magnetosphere size ($R_{\rm mp}$). This is done by considering magnetic and stellar wind dynamic pressure balance (e.g., \citeads{1969JGR....74.1275S}, \citeads{2004pse..book.....G}) leading to the relation: 

\begin{equation}\label{eq:1}
\dfrac{R_{\rm mp}}{R_{\rm p}} = \left(\dfrac{B_{\rm p}^2}{8\pi P_{\rm dyn }}\right)^{1/6}\mbox{\hspace{-0.3cm}.}
\end{equation}

\noindent Assigning an Earth-like dipole magnetic field $B_{\rm p}$\,$=$\,$0.3$~G to Proxima c, this calculation yields a magnetosphere size ranging between $\sim$\,$6-8~R_{\rm p}$ among both activity states (see~Fig.~\ref{fig:orbits}, bottom panels). These values appear close to the standard size of the day-side Earth's magnetosphere ($\sim$\,$10$~$R_{\rm earth}$), which can be compressed by up to $\sim$\,$35\%$ during strong solar space weather events (see~\citeads{2007LRSP....4....1P}, \citeads{2015GeoRL..42.4694L}). Figure~\ref{fig:orbits} also shows the inverse relation between the dynamic pressure and the magnetosphere size, with crossings of the current sheet as coinciding peaks and dips in $P_{\rm dyn}$ and $R_{\rm mp}$, respectively. As can be seen from Eq.~\ref{eq:1}, $R_{\rm mp} \propto B_{\rm p}^{1/3}$, so that larger magnetosphere sizes are expected for stronger planetary magnetic field values.

\begin{table*}[!htbp]
\caption{Representative stellar wind parameters around Proxima~c and resulting properties from the GM+IE modules.}             
\label{table_1}      
\centering
{\small        
\begin{tabular}{ l c c c c | c c c }    
\hline\hline
\\[-10pt]
\multirow{2}{*}{Case} &  \multicolumn{4}{c|}{Incident Stellar Wind} & \multicolumn{3}{c}{Global Magnetosphere (GM+IE)} \\
 & $n$ [cm$^{-3}$] & $T$ [$\times$\,10$^{4}$ K] & $\mathbf{U}$ [km s$^{-1}$] & $\mathbf{B}$ [nT] & $R^{\rm\,min}_{\rm mp}$ [$R_{\rm p}$] & $P^{\rm\,max}_{\rm mp}$ [$P_{\rm earth}$] & JH [JH$_{\,\rm earth}$] \\[2pt]
\hline
& & & & & & &\\[-9pt] 
Minimum & 1.0 & 5.0 & ($-1100, 0, 0$) & ($0, 0, -0.5$)  & 8.2 & 1.3 & 0.07\\
Maximum & 10.0 & 10.0 & ($-600, 0, 0$) & ($0, 0, -2.0$) & 6.2 & 2.4 & 0.16\\\\[-10pt]
\hline                  
\end{tabular}}
\end{table*}

\begin{figure*}
\center %  left, bottom, right, top
\includegraphics[trim = 0.0cm 0.0cm 0.0cm 0.cm, clip=true,width=0.495\textwidth]{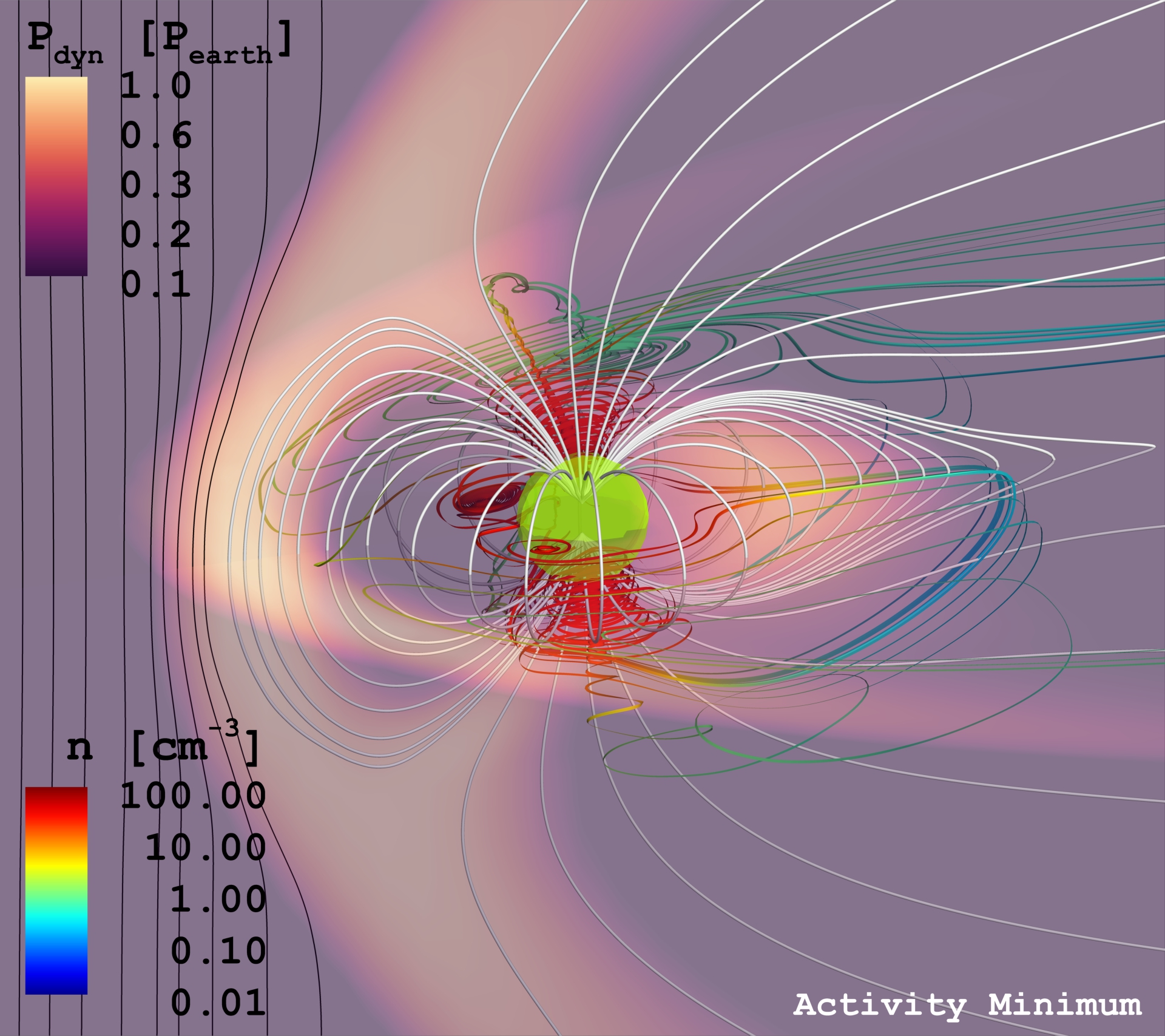}\hspace{1pt}\includegraphics[trim = 0.0cm 0.0cm 0.0cm 0.0cm, clip=true,width=0.495\textwidth]{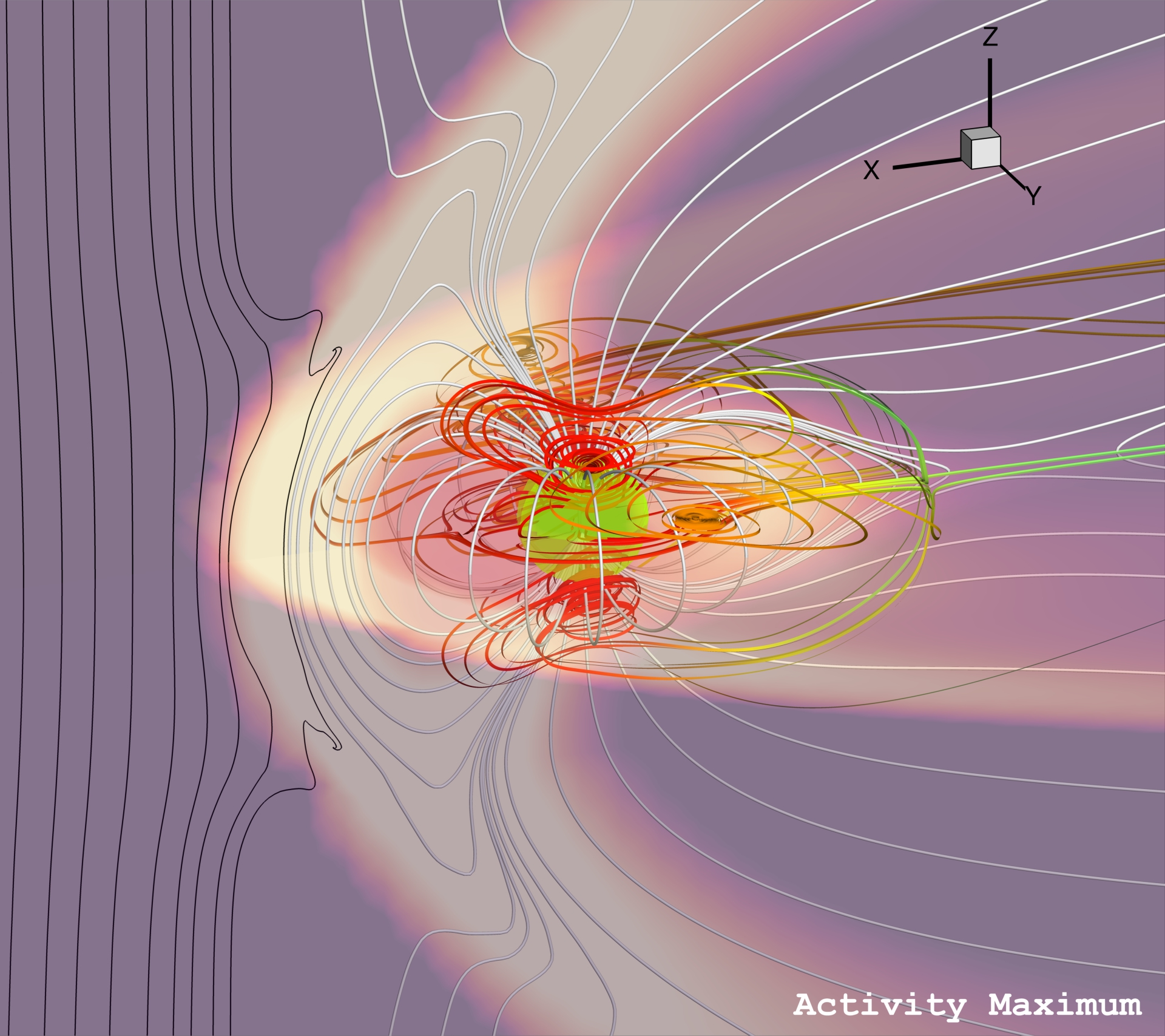}
\caption{Results from the GM+IE model driven by stellar wind parameters representative of both activity states (minimum:~left, maximum: right; see Table~\ref{table_1}), extracted from the analyzed orbits of Proxima c within the IH module (Figs.~\ref{fig_1} and \ref{fig:Pdyn_2D}). The star is located in the positive $x$ direction and the central sphere (green) corresponds to the inner boundary of the domain ($R = 2~R_{\rm p}$). Equatorial and meridional projections of the normalized wind dynamic pressure are included ($P_{\rm dyn}$, burgundy). Randomly seeded velocity streamlines, color coded by plasma number density ($n$, rainbow), are used as proxy for particle trajectories inside the magnetosphere. Selected stellar wind (black) and planetary (white) magnetic field lines are shown.}
\label{fig_GM}
\end{figure*}

To complement the analytic description, our multi-domain simulation also includes a three-dimensional model of a possible magnetosphere and ionosphere around Proxima~c (see~Sect.~\ref{sec:Methods}). We evaluate the stellar wind properties along the considered orbits of Proxima~c in order to obtain nominal conditions\,---namely density, speed, magnetic field strength, and temperature---\, in each activity state (see Table~\ref{table_1}). These representative stellar wind parameters are used to drive the GM and IE modules, whose results are presented in Fig.~\ref{fig_GM}. The visualizations include equatorial and meridional projections of $P_{\rm dyn}$, clearly showing the development of a bow shock towards the star (positive $x$-axis). As expected, the harsher stellar wind conditions during maximum generate higher compression of the entire magnetosphere compared to activity minimum. In combination with the polarity and strength of the interplanetary magnetic field, this will influence the fraction of particles penetrating and precipitating to the ionosphere (illustrated in Fig.~\ref{fig_GM} by density-colored velocity streamlines). A summary of the resulting values from the GM module is presented in Table~\ref{table_1}. Note that the smallest magnetosphere standoff distance from GM is fairly consistent with the analytic formulation given by Eq.~\ref{eq:1} (see also Fig.~\ref{fig:orbits}, lower panels). 

Following \citetads{2020ApJ...897..101C}, we calculate the associated Joule heating in the upper atmosphere using the IE model. We find that the Joule Heating is very low for both activity states\,---about $\sim$\,$10\%$ of the heating obtained at Earth during ambient solar wind conditions (JH$_{\,\rm earth} \simeq 150$~GW; see Table~\ref{table_1}). The reason for this is that while $P_{\rm dyn}$ is higher for Proxima~c than for the Earth, the average solar wind conditions carry a stronger magnetic field (particularly in the $B_{\rm z}$ component), whose variations ultimately drive the field aligned currents and the particle influx responsible for the JH. Previous studies of the TRAPPIST-1 planets \citepads{2018ApJ...856L..11C} and TOI-700~d \citepads{2020ApJ...897..101C} indicate much higher values of JH, that could potentially contribute to continuous heating of the upper atmosphere of these planets. However, such heating is likely negligible for Proxima~c.

\subsection{Extreme Conditions for Proxima~d}\label{sec:Prox-d}

\noindent To complete this study, we examine the expected space environment around the planet candidate Proxima~d. As mentioned in Sect.~\ref{s:proxima}, its $\sim$\,$40.4~R_{\bigstar}$ orbit would place it closer than Proxima~b ($a \simeq 67.8~R_{\bigstar}$), exposing it to even more extreme conditions than the habitable zone planet. This includes $P_{\rm dyn}$ values about $5$ times larger than expectations for Proxima~b\footnote[15]{Comparing with the $\left<B\right>_{\rm S} = 600$~G case from \citetads{2016ApJ...833L...4G}.} \citepads{2016ApJ...833L...4G}, corresponding to $3$--$4$ orders of magnitude larger than what the present day Earth experiences (see~Fig.~\ref{fig_1}, middle panels). While both exoplanets would face similar intra-orbital variations in $P_{\rm dyn}$ (by a factor of $\sim$\,$10$), they will occur approximately twice as fast in Proxima~d compared to b ($\sim$\,$5.2$~d versus $\sim$\,$11.2$~d orbital periods). Furthermore, the $46.5~R_{\bigstar}$ average size of the AS during maximum\,---larger than the orbital separation---\,implies that Proxima~d would cross (or be completely in) sub-Alfv\'enic stellar wind sectors at times of high-activity in the star. 
Leaving aside the increased rate of high-energy transients and their expected strong coronal response on Proxima (see~\citeads{2019ApJ...884L..13A}), the sub-Alfv\'enic conditions pose an even greater challenge for the retention of any atmosphere around the planet (e.g.~analogous to the case of the TRAPPIST-1 system; \citeads{2017ApJ...843L..33G}, \citeads{2018ApJ...856L..11C}).

\section{Summary and Conclusions}

\noindent As the closest planetary system to Earth, Proxima Centauri and its circumstellar properties are of great importance for exoplanet and habitability studies. To characterize the expected conditions of the recently discovered Proxima~c, we have constructed the most comprehensive numerical simulation of the space environment in this system to date. This includes coupled models for the stellar corona and inner astrosphere\,---where the complete $\sim$\,$1.44$~au orbit of the planet is enclosed---\,driven by realistic surface magnetic field configurations representative of the minimum and maximum activity states of Proxima.  

Our results indicate that Proxima~c experiences Earth-like conditions\,---in terms of the dynamic pressure exerted by the stellar wind---\,along its $\sim$\,$5.3$~yr orbit, with minor variability (by a factor of $\sim$\,$2$) due to the activity cycle of the star. To investigate the relative effect of such conditions on the energy dissipation in the upper atmosphere (Joule heating), we also simulated a possible magnetosphere and ionosphere around the planet. We found that even with a relatively weak planetary dipole field ($0.3$~G), the associated Joule heating of the upper atmosphere is negligible for Proxima~c ($\sim$\,$10$\% of the nominal value on the Earth), due to a diminished interplanetary magnetic field at the distance of the planet. Whether or not Proxima Cen c currently has an atmosphere would depend on several factors, including its formation channel and evolutionary path. 
Nevertheless, at face value the resulting conditions from our models do not appear to be unduly corrosive and should be favorable for the persistence of any extant atmosphere, supporting the prospect of fruitful future observing campaigns.  

Finally, we also examined the resulting space environment around the planet candidate Proxima~d, which is expected to orbit at only $0.029$~au. Not surprisingly, this exoplanet would experience extreme conditions, including very large dynamic pressures ($10^3 - 10^4$ times the average value around the Earth) with sub-orbital variability reaching factors of 10, and even the possibility of sub-Alfv\'enic conditions for extended periods of time. A grim space weather forecast is then expected for this exoplanet candidate.
\\

%% Putting eqnarrays or equations inside the mathletters environment groups
%% the enclosed equations by letter. For instance, the eqnarray below, instead
%% of being numbered, say, (4) and (5), would be numbered (4a) and (4b).
%% LaTeX the paper and look at the output to see the results.


\acknowledgments
\noindent We would like to thank the referee for constructive feedback. Support for Program number HST-GO-15326 was provided by NASA through a grant from the Space Telescope Science Institute, which is operated by the Association of Universities for Research in Astronomy, Incorporated, under NASA contract NAS5-26555. JJD was funded by NASA contract NAS8-03060 to the CXC and thanks the Director, Belinda Wilkes, for continuing advice and support. OC was supported by NASA NExSS grant NNX15AE05G. KP acknow\-led\-ges funding from the German \textit{Leibniz Gemeinschaft} under project number P67-2018. This work used SWMF/BATSRUS tools developed at The University of Michigan Center for Space Environment Modeling. Simulations were performed on NASA's Pleiades cluster under award SMD-17-1330, provided by the NASA High-End Computing Program through the NASA Advanced Supercomputing Division at Ames Research Center.


\facilities{Pleiades}


\software{SWMF \citepads{2018LRSP...15....4G}}

\end{document}